# ChatGPT Performance on Standardized Testing Exam –A Proposed Strategy for Learners

Umer Farooq
Ph.D. Student
Department of Multidisciplinary Engineering
Texas A&M University
College Station, TX, USA
umerfarooq@tamu.edu

Saira Anwar
Assistant Professor
Department of Multidisciplinary Engineering
Texas A&M University
College Station, TX, USA
sairaanwar@tamu.edu

*Abstract*— **This study explores the problem-solving capabilities of ChatGPT and its prospective applications in standardized test preparation, focusing on the GRE quantitative exam. Prior research has shown great potential for the utilization of ChatGPT for academic purposes in revolutionizing the approach to studying across various disciplines. We investigate how ChatGPT performs across various question types in the GRE quantitative domain, and how modifying question prompts impacts its accuracy. More specifically this study addressed two research questions: 1. How does ChatGPT perform in answering GRE-based quantitative questions across various content areas? 2. How does the accuracy of ChatGPT vary with modifying the question prompts? The dataset consisting of 100 randomly selected GRE quantitative questions was collected from the ETS official guide to GRE test preparation. We used quantitative evaluation to answer our first research question, and t-test to examine the statistical association between prompt modification and ChatGPT's accuracy. Results show a statistical improvement in the ChatGPT's accuracy after applying instruction priming and contextual prompts to the original questions. ChatGPT showed 84% accuracy with the modified prompts compared to 69% with the original data. The study discusses the areas where ChatGPT struggled with certain questions and how modifications can be helpful for preparing for standardized tests like GRE and provides future directions for prompt modifications.**

I. INTRODUCTION

Graduate Record Examination (GRE) plays a significant role in evaluating and comparing applicants' cognitive abilities for graduate studies [1, 2]. Every year, thousands of students worldwide take the GRE as part of their application process for graduate programs. The GRE test comprises three main sections: verbal reasoning, quantitative reasoning, and analytical writing.

The Quantitative Reasoning section of the GRE assesses students' aptitude to tackle challenging problems by applying fundamental mathematical concepts and doing data analysis. The preparation of this section is not only time-consuming [3] but requires considerable expertise. Students frequently struggle in solving various questions and may seek expert assistance to comprehend complex concepts. Moreover, access to comprehensive and freely available resources can be limited, which adds to the challenges faced by students during their GRE preparation journey.

As Artificial Intelligence (AI) technology, particularly in the domain of Natural Language Processing (NLP), continues to advance, users can now interact with more natural and intuitive chatbots [4]. ChatGPT, a sophisticated chatbot technology that combines AI and NLP [5] can be a valuable tool for students preparing their GRE exam's quantitative section. It offers great potential in revolutionizing the approach to studying and getting ready for the GRE. ChatGPT offers the convenience of providing accurate solutions to specific problems, thereby streamlining the process, and facilitating better preparation.

This study aims to explore ChatGPT's problem-solving abilities and potential applications in educational settings. To evaluate ChatGPT's performance, we investigated how ChatGPT handles GRE quantitative questions across various question types. Furthermore, how students can modify their question prompts to maximize the potential use of ChatGPT in their GRE quantitative exam preparation. Specifically, in this study, we answered the following research questions.

1. How does ChatGPT perform in answering GRE-based quantitative questions across various content areas?

2. How does the accuracy of ChatGPT vary with modifying the question prompts?

By evaluating the performance of ChatGPT on GRE quantitative problems, we are contributing to the advancement of AI-powered educational tools and platforms. The findings of this study will enhance our understanding of the potential use and effectiveness of ChatGPT in educational settings, specifically in the realm of standardized test preparation, and how we can modify prompts to maximize its performance.

The remaining sections of the paper are structured as follows. Section II provides a comprehensive review of the existing literature on the use of AI and NLP in educational settings, highlighting the advantages they offer as well as their limitations. Section III presents the methodology employed in this study, including the data collection process and the specific tasks performed using ChatGPT. Section IV presents the analysis and results, with a focus on the performance and effectiveness of ChatGPT in providing answers. Section V offers a detailed discussion of the findings, addressing their implications for educational practice and potential future research directions. Finally, Section VI concludes the paper by summarizing the key findings and the significance of integrating AI chatbots like ChatGPT in educational settings.

## II. LITERATURE REVIEW

With growing research on artificial intelligence and language models, many studies have discussed the effectiveness of ChatGPT [6, 7]. ChatGPT is an advanced OpenAI software, functioning as a large language model. It is developed by OpenAI (chat.openai.com). ChatGPT 3.5 is available to the general public with an aim to enhance natural language interactions [8].

Numerous studies have been conducted to evaluate the abilities of language models such as ChatGPT. These prior studies on the effectiveness of ChatGPT can be classified into three major categories: 1) Language and literature studies [9], 2) Scientific and technical studies [10], and 3) Social sciences and knowledge dissemination [11].

In the realm of language and literature studies, researchers examined ChatGPT's ability to produce coherent and contextually appropriate texts. Iyolita Islam, Muhammad Nazrul Islam explored the prospective applications of ChatGPT in academic writing [12]. Their study highlighted ChatGPT's ability to generate contextually appropriate essays. ChatGPT exhibited proficiency while adapting to different writing styles and provided feedback to any given text. Additionally, ChatGPT's ability to provide real-time answers and support to multiple languages further expands its applicability to a wide range of multidisciplinary tasks [13].

For the scientific and technical studies, researchers accessed ChatGPT's ability to process given scientific questions ranging from medicines, engineering, and sciences majors. For example, X. Hu et al [14] focused on the potential use of ChatGPT to facilitate design knowledge acquisition. Their findings revealed that ChatGPT provided more accurate and comprehensive information, which enhanced the integration of generated knowledge into the design process. Kung, Tiffany H., et al. evaluated ChatGPT on its performance in the United State Medical Licensing Exam (USMLE) [10]. Their findings suggest that ChatGPT can pass the exam without any specialized training.

In the domain of social sciences and knowledge dissemination, researchers investigated how ChatGPT contributed to answering factual questions and generating educational content. E. Kasneci et al. highlighted how ChatGPT can be used for not only creating educational content but also to improve student engagement and a personalized learning tool [15]. Junaid Qadir explored the potential of ChatGPT for engineering education [16]. He highlighted that ChatGPT can not only offer personalized experience to student but a more effective one as it provides students with customized feedback and explanations.

Given the capabilities of ChatGPT, there are certain limitations to using it as well. These include the potential for biased outputs [17], limited abilities to answer scientific research questions requiring source information [12], and cheating and plagiarism [18]. To integrate large language models like ChatGPT into the educational domain, it is an absolute necessity for both instructors and students to develop extensive skills [15]. This means that not only a profound understanding of how to use ChatGPT for preparing GRE exam but also acknowledging the weaknesses and vulnerabilities and way of improving the prompts in such large language models. This paper is focusing on the aspect of using ChatGPT for preparing GRE quantitative exam and providing a way out for maximizing the effectiveness in the educational domain.

## III. RESEARCH DESIGN

The research conducted in this study adheres to a quantitative methodology. Data collection occurred at a specific moment in time (Early June 2023). We opted for a quantitative evaluation research design to address our first research question. The evaluation research design enables a comparison between ChatGPT's performance and a benchmark. We used the Educational Testing Service (ETS) Official GRE General Test Preparation book as a benchmark for the GRE quantitative exam. By comparing ChatGPT's answers to the benchmark, we assessed its accuracy and identified areas for improvement. We chose quantitative research design to examine the statistical association between prompt modification and ChatGPT's accuracy in the second research question. We conducted a t-test to explore statistical differences in the accuracy of ChatGPT after we modified the prompts.

### A. Measures and Data Collection

We collected the data from the ETS official guide to the GRE general test preparation 3rd edition book. ETS which administers the GRE, provides a pool of questions that can be used for preparation. These questions are designed to cover various topics and difficulty levels that are representative of the actual GRE exam.

A total of 100 randomly selected GRE quantitative questions were chosen for this study, assuming that all question examples ChatGPT model 3.5 had not been encountered during its training. The selected questions went through a screening process to remove any items that included visuals such as graphs, tables, and geometry. Following this filtering process, the final set of questions included 43 Numeric Entries, 22 Quantitative Comparisons, 18 multiple-choice questions with a single correct answer, and 17 multiple-choice questions with multiple correct answers. Table 1 provides an overview of the distribution of question types within different question categories.

TABLE 1. QUESTIONS CATEGORY WITH CORRESPONDING NUMBER OF QUESTION

| Question Types | Total |
| --- | --- |
| Numeric Entry | 43 |
| Quantitative Comparisons | 22 |
| Multiple Choice (Single answer) | 18 |
| Multiple Choice (Multiple answers) | 17 |

| Total | 100 |
|---|---|

We observed that ChatGPT was finding difficulties with the questions involving ambiguous relational operators. To maximize the utility of ChatGPT, we made modifications to the prompts of such questions. By incorporating specific context within each line and additional information about the relational statement to each question, we observed improvements in ChatGPT's answers.

*B. Prompt Modification*

To enhance the capabilities of ChatGPT for solving GRE quantitative questions, we utilized instruction priming [19] and contextual prompts [20] as prompt modification strategies. Table 2 presents a selection of original questions alongside their modified versions, illustrating the variations in ChatGPT's responses before and after prompt modification.

1. Instruction Priming

Instruction priming is a preparatory strategy that familiarizes students with upcoming information or learning activities in a course, enhancing their academic learning [21]. For the ChatGPT prompts, we introduced specific topics before engaging with it. This preparatory approach enhanced the quality and relevance of ChatGPT responses. Instruction priming ensured that ChatGPT is better prepared to respond to GRE quantitative questions. For example, since ChatGPT was unable to calculate interquartile range correctly, we primed ChatGPT with the correct mathematical approach first and then asked it to solve the question.

2. Contextual Prompts

Contextual prompts are prompts tailored to the user's specific context or scenarios. They provide suggestions aligned with the current situation or needs. For the ChatGPT prompts, contextual prompts helped clarifying ambiguous queries. By adding relevant and contextual information, ChatGPT comprehended the query in better way and stopped generating irrelevant and confusing responses.

TABLE 2. Sample Original Vs Modified Prompts

| \multicolumn{3}{c}{**Type of Question**} |
|---|---|---|
| \multicolumn{3}{c}{*"Quantitative Comparison"*} |
| Original Prompt | Set S consists of all positive integers less than 81 that are not equal to the square of an integer.<br><br>Quantity A: The number of integers in set S<br>Quantity B:   72 | Wrong answer due to misinterpretation of the square of an integer. |
| Modified Prompt | Consider set S, which consists of all positive integers less than 81 that are not perfect squares.<br><br>Quantity A: The number of integers in set S<br>Quantity B:   72 | Correct answer due to adding information of perfect square (context). |
| \multicolumn{3}{c}{**Type of Question**} |
| \multicolumn{3}{c}{*"Multiple Choice Question"*} |
| Original Prompt | A certain shipping service charges an insurance fee of $0.75 when shipping any package with contents worth $25.00 or less and an insurance fee of $1.00 when shipping any package with contents worth over $25.00. If Dan uses the shipping company to ship three packages with contents worth $18.25, $25.00, and $127.50, respectively, what is the total insurance fee that the company charges Dan to ship the three packages?<br>A) $1.75<br>B) $2.25<br>C) $2.50<br>D) $2.75<br>E) $3.00 | Wrong answer due to ambiguity related to the relational operator present in the question. |
| Modified Prompt | A certain shipping service charges an insurance fee of $0.75 when shipping any package valued equal to $25.00 or less. For packages valued over $25.00, the insurance fee increases to $1.00. If Dan uses the shipping company to ship three packages with contents worth $18.25, $25.00, and $127.50, respectively, what is the total insurance fee that the company charges Dan to ship the three packages?<br>A) $1.75<br>B) $2.25<br>C) $2.50 | Correct answer due to clarification of the package value (more explanation relevant to the relational operator). |

| | D) $2.75<br>E) $3.00 | |
| --- | --- | --- |
| **Type of Question**<br>*"Numeric Entry"* | | |
| Original Prompt | How many 3-digit positive integers are odd and do not contain the digit 5? | Wrong answer due to ambiguity regarding digit 5 location. |
| Modifies Prompt | How many 3-digit positive integers are odd and do not contain the digit 5? The first digit can be any non-zero integer but 5, and the third digit can be any odd number except 5. | Correct answer due to elaboration within each line (priming). |

*C. Procedure and Analysis*

To get the answers from The ChatGPT model, we provided ChatGPT with each question in the form of a prompt or query. ChatGPT generated a response, offering its explanation and answer to the provided question. We recorded the answers generated by ChatGPT, along with their corresponding explanation, to facilitate subsequent analysis.

For the first research question, we compared the answers generated by ChatGPT to the benchmark answers obtained from the ETS official test preparation book to assess their accuracy. We determined the accuracy rate of ChatGPT by calculating the percentage of questions that were answered correctly and matched with the answers provided in the benchmark. Further, we examined ChatGPT's performance variations across question categories, by computing the accuracy rates individually for each question type, namely Numeric Entry, Quantitative Comparison, and Multiple Choice.

For the second research question, we presented ChatGPT modified question prompts for the ones it attempted wrong on the first attempt. For prompt modification, we added context related to the problem by elaborating on the details. By incorporating relevant contextual information and applying specific constraints such as requesting to use specific mathematic rules, asking to adhere to certain rules while answering, and providing explicit instructions, we tested ChatGPT's performance.

After prompt modifications, we again compared ChatGPT's generated answers to the benchmark answers to determine the accuracy. To further explore the relationship between prompt modification and ChatGPT's accuracy, we conducted an independent sample t-test. We aimed to compare mean scores of ChatGPT's accuracies based on prompt modification. The alpha level was set at 0.05 to determine the significance. The t-test was performed using SPSS v 29.0 P-value was automatically calculated by the software for the independent sample t-test.

## IV. FINDINGS

To answer our first research question, we provided ChatGPT with 100 questions and compared the answers with the corresponding answers available in the benchmark book. Dividing the number of correct answers provided by ChatGPT with the total number of questions, we calculated the accuracy. Table 3 indicates a summary of ChatGPT's performance.

TABLE 3. ChatGPT Performance on GRE Quantitative Questions

| Total number of questions | Correctly attempted questions | Accuracy |
| --- | --- | --- |
| 100 | 69 | 69% |

The accuracy value indicates how ChatGPT performed when we provided it with GRE quantitative questions. The GRE quantitative section consists of different question types which are not proposed in this research but rather categorized by ETS. These question types include numeric entry, quantitative comparison, multiple choice with one correct answer, and multiple choice with one or multiple correct answers. In numeric entry questions, there is a single numeric answer which can be an integer or fraction. For quantitative comparison-based questions, with every statement, there are two statements associated to compare against each other. The answer could either be statement A is greater or statement B is greater, or both are equal, or the comparison is not possible with the given information. For Multiple choice single correct-answer questions, there is only one correct option with every question. Whereas, for Multiple choice multiple correct-answer questions, there can be one correct option or multiple and to do it correctly every correct option needs to be chosen.

We split our 100-question dataset into these four categories to gain deeper insight into ChatGPT's performance in each question type. Table 4 indicates a summary of ChatGPT's performance against each question type.

TABLE 4. ChatGPT performance against different question types

| Question type | Total number of questions | Correctly attempted questions | Accuracy |
| --- | --- | --- | --- |
| Numeric Entry | 43 | 37 | 86.04% |
| Quantitative Comparison | 22 | 12 | 54.54% |
| Multiple Choice (Single Answer) | 18 | 13 | 72.22% |
| Multiple Choice (Multiple Answers) | 17 | 07 | 41.17% |

When provided with numeric entry questions, we observed the highest accuracy from ChatGPT followed by Multiple-choice questions requiring one single answer. Questions requiring quantitative comparisons posed certain challenges to ChatGPT and ranked third in terms of accuracy. ChatGPT did not perform well on the multiple-choice questions having multiple correct options. The accuracy comparison among various question types is further depicted in fig 1.

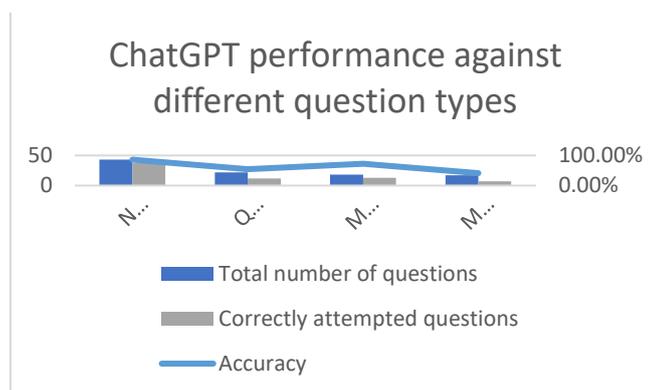

Fig1: ChatGPT performance against different question types

To address our second research question, we provided ChatGPT with 31 questions that were initially attempted incorrectly. We then modified the question prompts using instruction priming and contextual prompts and compared the ChatGPT responses with the corresponding answers from the benchmark book. To calculate the accuracy, we divided the total number of correct answers provided by ChatGPT before and after re-prompting by the total number of questions. Table 5 provides an overview of ChatGPT's performance after we modified the prompts.

TABLE 5. ChatGPT performance on GRE Quantitative questions

| Total number of questions | Correctly attempted questions | Accuracy |
|---|---|---|
| 100 | 84 | 84.0% |

To check for statistical significance, we conducted independent comparisons to gauge the efficacy of ChatGPT under modified prompts compared to the original questions, using an independent sample t-test. For modified prompts, Levene's test (F = 26.902, p < 0.001) was significant; therefore, we used the t-test for equal variance not assumed. The results highlight that ChatGPT performance was different when given the original questions and after we modify the prompts (see Table 6).

The results showed a statistically significant difference between both groups, with adjusted degrees of freedom t (188.195) = -2.529, p = 0.012. Cohen's d = 0.41942, which showed a medium effect size [22]. These results indicate that ChatGPT performed better with the questions that were modified than the original questions.

TABLE 6: Difference Between the Performance of ChatGPT with Original and Modified Prompts

| Original N = 100 | | Modified | | | | | |
|---|---|---|---|---|---|---|---|
| M | SD | M | SD | df | t | p | Cohen's d |
| 0.690 | 0.464 | 0.840 | 0.368 | 188.195 | -2.529 | 0.012 | 0.419 |

We encompassed a diverse set of question prompt types, including numerical problems requiring relational comparisons, statistical analysis, computing data ranges and interquartile ranges, and multiple-choice questions. We found out that the accuracy of ChatGPT's responses was influenced by the nature of the prompt or question type. Specifically, when we tried questions related to interquartile range calculations, finding the mean, median, and mode of data, and interpreting complex relational operations in quantitative reasoning questions, ChatGPT often provided inaccurate results.

However, we observed clear improvement in ChatGPT's responses when we primed it with relevant information in such questions. By specifying the mathematical process required to solve interquartile range questions, as well as explaining slightly implicit details in quantitative reasoning questions.

We run the individual tests as well on the four categories of GRE quantitative questions. Since the number of samples in each category except Numeric Entry was less than 30, we did not use the traditional parametric tests. Instead, we worked with Mann-Whitney U-test which is a non-parametric test and does not require assumptions related to data distribution. The results highlight that ChatGPT performance was better with quantitative comparisons after we modified the prompts. (see Table 7).

TABLE 7: Difference Between the Performance of ChatGPT with Original and Modified Prompts

|  | Original | | Modified | | | |
| --- | --- | --- | --- | --- | --- | --- |
|  | M | SD | M | SD | U | p |
| Numeric Entry | 0.895 | 0.307 | 1.50 | 0.502 | 860 | 0.29 |
| Quantitative Comparisons | 0.681 | 0.471 | 1.50 | 0.505 | 176 | 0.05 |
| MCQ (Single correct) | 0.805 | 0.401 | 1.50 | 0.507 | 135 | 0.21 |
| MCQ (Multiple correct) | 0.500 | 0.507 | 1.50 | 0.507 | 119 | 0.31 |

## V. Discussion

This study used ChatGPT model 3.5 and explored its accuracy for various question types of GRE quantitative exams. These question types included numeric entry, quantitative comparisons, multiple-choice questions having a single correct answer, and multiple-choice questions having multiple correct answers. Additionally, the study investigated the improvements of modifying certain questions for the overall ChatGPT accuracy. The study provides an intriguing insight into how well ChatGPT attempted GRE quantitative questions. The result of the first research question indicated that ChatGPT achieved an accuracy of 69% on a blinded dataset. Out of 100 randomly selected questions, it provided both the correct answers and explanations for 69 questions. This accuracy suggests that ChatGPT can be a useful tool for students preparing for the GRE exam. However, it is important to note that relying solely on ChatGPT may not be advisable.

The findings from the second research question provided a considerable improvement in ChatGPT's accuracy. We observed that when additional details were incorporated into the question prompts, ChatGPT was able to provide more concrete answers. Specifically, quantitative comparisons and questions involving determining the mean, median, and mode of given data, as well as calculating the interquartile range, posed challenges for ChatGPT. Regarding interquartile range questions, ChatGPT's approach of considering the median value to determine Q1 was found to be mathematically incorrect. Likewise, in quantitative comparison questions, ChatGPT failed to identify certain implicit logical comparisons within the questions.

Nevertheless, the performance of ChatGPT showed improvement when incorporated more explicit small checks and specifying the correct mathematical procedures for solving various problems. These improvements in question prompts positively impacted ChatGPT's accuracy, highlighting the importance of offering sufficient context and clear instructions to utilize ChatGPT in a better way for GRE test preparation. T-test result also showed the statistical significance of adding context to the question prompts.

The findings of this study should be interpreted with consideration of several limitations that may impact their applicability and generalization. Firstly, the sample size used might not fully represent the diversity of questions encountered in actual tests, potentially limiting the study's ability to account for all possible scenarios. Secondly, although prompt modifications were shown to enhance ChatGPT's accuracy, the most effective modifications might vary depending on the specific question content and domain. Lastly, it's essential to acknowledge that ChatGPT is an AI language model that continuously learns and evolves with each question prompt. Therefore, the study's results pertain to a specific point in time and may not necessarily extend to future iterations of the model. Moving forward, future research could identify the most effective prompt modifications.

## VI. Conclusion

In conclusion, this research contributes to our comprehension of ChatGPT's capabilities and provides valuable insights into its utilization for educational purposes. The outcomes of this study lead to improved interactions with ChatGPT. The results revealed that ChatGPT exhibited a promising level of accuracy in answering GRE-based quantitative questions, achieving an overall accuracy rate of 73.56% on a blinded dataset. However, the study also identified certain limitations in ChatGPT's performance. Specifically, the model encountered challenges in tasks involving interquartile range calculations, quantitative comparisons, and interpretation of complex relational operations. Prompt modifications played a crucial role in enhancing accuracy and to yield more accurate results.

## References

[1] D. F. Feldon, K. Litson, B. Cahoon, Z. Feng, A. Walker, and C. Tofel-Grehl, "The predictive validity of the GRE across graduate outcomes: A meta-analysis of trends over time," The Journal of Higher Education, pp. 1-29, 2023.

[2] A. Bleske-Rechek and K. Browne, "Trends in GRE scores and graduate enrollments by gender and ethnicity," Intelligence, vol. 46, pp. 25-34, 2014.

[3] J. M. Miller, A. Goodyear-Orwat, and R. W. Malott, "The effects of intensive, extensive, structured study on GRE scores," Journal of Behavioral Education, pp. 369-379, 1996.

[4] M. Firat, "How chat GPT can transform autodidactic experiences and open education," Department of Distance Education, Open Education Faculty, Anadolu Unive, 2023.

[5] S. S. Biswas, "Potential use of chat gpt in global warming," Annals of biomedical engineering, vol. 51, no. 6, pp. 1126-1127, 2023.

[6] D. Baidoo-Anu and L. Owusu Ansah, "Education in the era of generative artificial intelligence (AI): Understanding the potential benefits of ChatGPT in promoting teaching and learning," Available at SSRN 4337484, 2023.

[7] J. Deng and Y. Lin, "The benefits and challenges of ChatGPT: An overview," Frontiers in Computing and Intelligent Systems, vol. 2, no. 2, pp. 81-83, 2022.

[8] A. Stojanov, "Learning with ChatGPT 3.5 as a more knowledgeable other: an autoethnographic study," International Journal of Educational Technology in Higher Education, vol. 20, no. 1, p. 35, 2023.

[9] D. Yan, "Impact of ChatGPT on learners in a L2 writing practicum: An exploratory investigation," Education and Information Technologies, pp. 1-25, 2023.

[10] T. H. Kung et al., "Performance of ChatGPT on USMLE: Potential for AI-assisted medical education using large language models," PLoS digital health, vol. 2, no. 2, p. e0000198, 2023.

[11] H. Gimpel et al., "Unlocking the power of generative AI models and systems such as GPT-4 and ChatGPT for higher education: A guide for students and lecturers," Hohenheim Discussion Papers in Business, Economics and Social Sciences, 2023.

[12] I. Islam and M. N. Islam, "Opportunities and challenges of chatgpt in academia: A conceptual analysis," Authorea Preprints, 2023.

[13] H. Singh and A. Singh, "ChatGPT: Systematic Review, Applications, and Agenda for Multidisciplinary Research," Journal of Chinese Economic and Business Studies, vol. 21, no. 2, pp. 193-212, 2023.

[14] X. Hu, Y. Tian, K. Nagato, M. Nakao, and A. Liu, "Opportunities and challenges of ChatGPT for design knowledge management," arXiv preprint arXiv:2304.02796, 2023.

[15] E. Kasneci et al., "ChatGPT for good? On opportunities and challenges of large language models for education," Learning and Individual Differences, vol. 103, p. 102274, 2023.

[16] J. Qadir, "Engineering education in the era of ChatGPT: Promise and pitfalls of generative AI for education," in 2023 IEEE Global Engineering Education Conference (EDUCON), 2023: IEEE, pp. 1-9.

[17] E. Ferrara, "Should chatgpt be biased? challenges and risks of bias in large language models," arXiv preprint arXiv:2304.03738, 2023.

[18] C. K. Lo, "What is the impact of ChatGPT on education? A rapid review of the literature," Education Sciences, vol. 13, no. 4, p. 410, 2023.

[19] A. K. Jitendra et al., "Teaching mathematical word problem solving: The quality of evidence for strategy instruction priming the problem structure," Journal of Learning Disabilities, vol. 48, no. 1, pp. 51-72, 2015.

[20] K. Mao, Z. Dou, H. Chen, F. Mo, and H. Qian, "Large Language Models Know Your Contextual Search Intent: A Prompting Framework for Conversational Search," arXiv preprint arXiv:2303.06573, 2023.

[21] C. C. Griffin, J. C. Gagnon, M. H. Jossi, T. G. Ulrich, and J. A. Myers, "Priming mathematics word problem structures in a rural elementary classroom," Rural Special Education Quarterly, vol. 37, no. 3, pp. 150-163, 2018.

[22] J. Cohen, "Statistical power analysis," 1988.